\documentclass{PoS}
\title{Search for dark matter annihilation in the center of the Earth with 8 years of IceCube data}

\ShortTitle{8-years Earth WIMP analysis}

\author{
The IceCube Collaboration\footnote{For collaboration list, see PoS(ICRC2019) 1177.}\\
{\itshape \href{http://icecube.wisc.edu/collaboration/authors/icrc19_icecube}{http://icecube.wisc.edu/collaboration/authors/icrc19\_icecube}}\\
E-mail: \email{giovanni.renzi@icecube.wisc.edu}
}

\abstract{
Dark matter particles in the galactic halo can scatter off particles in celestial bodies such as stars or planets, lose energy and become gravitationally trapped. In this process, an accumulation of dark matter in the center of celestial bodies is expected, for example, at the center of the Earth. If dark matter self-annihilates into Standard Model particles, the end products of these annihilations include neutrinos. The IceCube Neutrino Observatory at the geographic South Pole can detect the resulting flux of neutrinos originating from dark matter annihilation in the center of the Earth. A search for this signal is on-going using 8 years of IceCube data and probing different annihilation channels. Here the sensitivities are presented for this new analysis, showing significant improvements with respect to the previous analyses from IceCube and other experiments. \\

\vspace{4mm}
{\bfseries Corresponding authors:}
\speaker{Giovanni Renzi}$^{1}$\\
{$^{1}$ \itshape Universit\'e Libre de Bruxelles}\\
}

\FullConference{36th International Cosmic Ray Conference -ICRC2019-\\
		July 24th - August 1st, 2019\\
		Madison, WI, U.S.A.}

\begin{document}

\section{Introduction}
\label{sec:intro}
Astronomical observations in the last century indicate the existence of a mass component in the Universe much larger than the contribution from baryonic matter only. This so-called Dark Matter (DM) has no clear physical explanation yet and several candidates have been proposed \cite{Bertone:2005} over the years. For particle DM, one of the most discussed candidates are the Weakly Interactive Massive Particles (WIMPs), predicted by supersymmetric extensions of the Standard Model (SM). In these models, WIMPs can scatter off SM particles in heavy celestial objects, such as Earth, lose energy and accumulate in the center of these bodies. Subsequently, WIMPs will self-annihilate into SM particles at a rate that is proportional to the DM density. Neutrinos are among the possible final products of these interactions and their resulting flux depends on the WIMP mass and annihilation channel.

IceCube has published upper limits on the spin-independent DM-nucleon scattering cross-section using 1 year of data \cite{EarthIce:2016}. The obtained results were competitive as compared to other experiments. In this work, an updated analysis, searching for an excess of neutrinos from the center of the Earth in 8 years of IceCube data, is presented.

\section{The IceCube Neutrino Telescope}
IceCube \cite{IceCube:2016a} is a cubic kilometer neutrino detector located at the geographic South Pole and installed in the Antarctic ice between depths of 1450 m and 2450 m. The detector consists of a large array of photomultipliers (PMTs) housed in glass spheres called Digital Optical Modules (DOMs). IceCube is composed of 86 vertical strings with 60 DOMs each and vertical spacing of 125 m. 
The DOMs record Cherenkov light emitted along the path of relativistic charged particles produced by neutrino interactions.
%When neutrinos interact producing charged particles, the passage of the latter in the detector volume causes the emission of Cherenkov light, which the DOMs record. 
The collected light allows reconstructing the characteristics of the primary neutrinos such as energy and direction. %The ice above the detector and the Earth shield down-going atmospheric muons and up-going atmospheric muons respectively; the former only below $\sim 500$ GeV.
Inside the IceCube volume, a smaller and denser array at a depth of 1750 cm, called DeepCore, is also installed. It consists of 8 closely-spaced strings in the center of the primary array with average sensor spacing of 72 m. DeepCore can use the remaining instrumented volume as a veto against muon and neutrino events originating from atmospheric interactions. DeepCore is of particular importance to detect neutrinos with energy below 100 GeV.% when considering low WIMP masses. The signal for this analysis consists of vertical up-going neutrino-induced muon tracks.

\section{Neutrinos from dark matter annihilation in the center of the Earth}
\label{sec:theory}
In order to be gravitationally captured in the center of the Earth, DM particles need to lose their initial kinetic energy by scattering off the matter nucleons in the planet. Given the relative abundances of Earth's chemical composition, this process is led by the spin-independent DM-nucleon scattering cross-section $\sigma_{\rm SI}$. The capture rate $C$ depends further on the DM mass, and the velocity and local density of DM particles at the position of the Earth. If the capture rate is constant in time, the annihilation rate is given by \cite{Bruch:2009}:
\begin{equation}
	\label{eq:ann_rate}
	\Gamma_A=\frac{C}{2}\tanh^2\left(\frac{t_{\oplus}}{\tau}\right)\mbox{ ,}
\end{equation}
where $\tau=(C\cdot C_A)^{-1/2}$ is the time scale for the capture and annihilation processes to reach equilibrium, the quantity $C_A$ describes annihilation and is proportional to the DM annihilation cross-section $<\sigma_A v>$, and $t_{\oplus} = 4.5\cdot10^9$ yrs is the age of Earth.

\begin{figure}
	\centering
	\includegraphics[width=.6\textwidth]{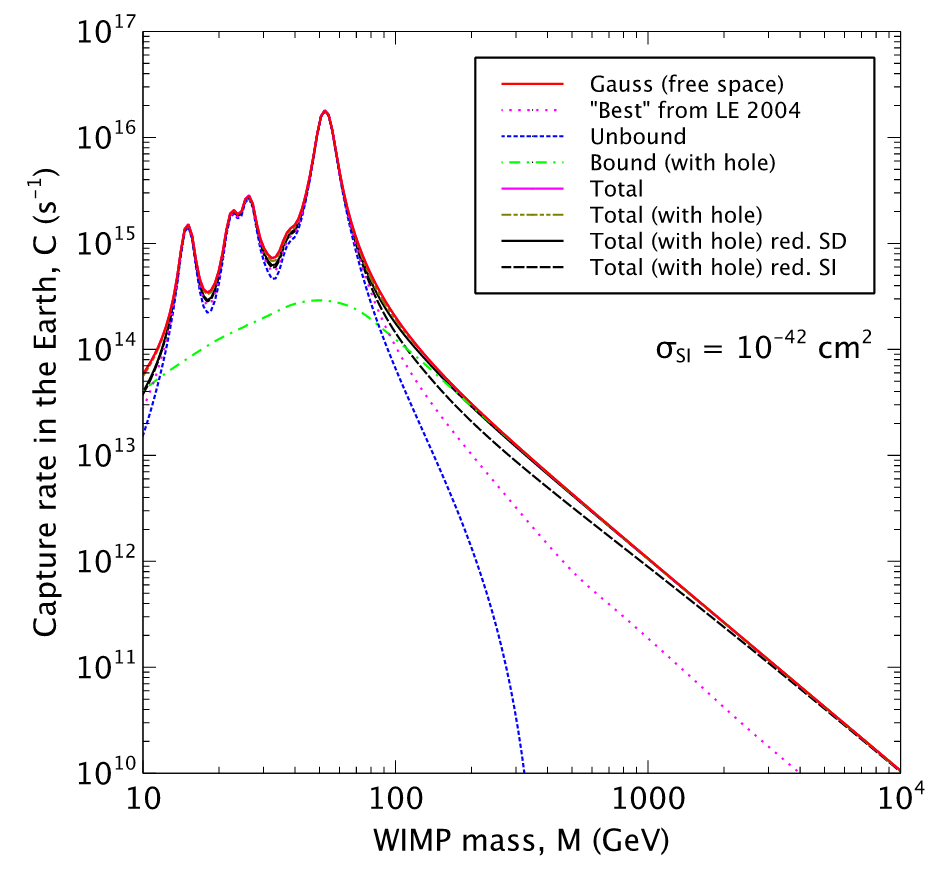}
	\caption{Capture rate of DM particles at Earth assuming $\sigma_{\rm SI}=10^{-42}$ cm$^2$. From \cite{Sivertsson:2012}.}
	\label{fig:capture_rate}	
\end{figure}

In the Standard Halo Model (SHM), the velocity distribution of DM particles in the galactic halo is assumed to follow a truncated Maxwellian distribution with a dispersion of $270$ km/s and an escape velocity of $544$ km/s \cite{DM_rev:96}. The SHM is adopted for this work and the local density is assumed to be \mbox{$0.3$ GeV/cm$^3$} for consistency with other experiments while its actual value is still under debate and can vary from $\sim0.2$ GeV/cm$^3$ to $\sim0.5$ GeV/cm$^3$ \cite{Read:2014}.
% DaM particles can have a resonant interaction with Earth atoms when their masses have very similar values.
Fig. \ref{fig:capture_rate} shows the capture rate in Earth as a function of the DM particle mass for a given value of $\sigma_{\rm SI}=10^{-42}$ cm$^2$.

The neutrino flux arising from DM annihilation in the center of the Earth is given by
\begin{equation}
	\label{eq:nu_flux}
	\frac{\textrm{d}\Phi}{\textrm{d}E_{\nu}}=\frac{\Gamma_A}{4\pi R^2_\oplus}\frac{\textrm{d}N_{\nu}}{\textrm{d}E_{\nu}}\mbox{ ,}
\end{equation}
where $R_{\oplus}$ is the Earth radius and  $\textrm{d}N_{\nu}/\textrm{d}E_{\nu}$ denotes the energy spectrum of secondary neutrinos produced in these annihilations, which depends on the WIMP mass and annihilation channel.

\section{Data and simulations}
%\subsection{Datasets}
In this work 8 years of experimental data  recorded between 2011 and 2018 are used. The background for this study are muons and neutrinos originating from cosmic-ray air-showers. Even though these muons can only have a down-going direction, a part of them is mis-reconstructed as up-going, constituting the main background for this analysis. Monte Carlo (MC) simulations are used to estimate the described background fluxes. Muons resulting from cosmic-ray air-showers are simulated with the CORSIKA software package \cite{corsika:98}. The Neutrino Generator (NuGen) and GENIE software packages are used to simulate neutrinos. Neutrino oscillations inside the Earth are taken into account for neutrino energies below \mbox{100 GeV} \cite{OscIce:2017}. A small subset of the data, corresponding to about 10\% of the total is used to validate the data/MC agreement in the signal region. This subset is disregarded for further analysis.

The neutrino signal, expected from DM annihilation, is simulated with the WimpSim \cite{WimpSim:2008} software package for various annihilation channels and DM masses. Two benchmark DM masses, $m_{\chi}=50$ GeV (\textit{low mass}) and $m_{\chi}=1$ TeV (\textit{high mass}), are used to validate the event selection for the annihilation channels $\chi\chi \rightarrow \tau^+\tau^-$ and $\chi\chi \rightarrow W^+W^-$ respectively. During the development of the event selection is was verified that the performance was comparable when considering other annihilation channels, as for example, $\chi\chi \rightarrow b\bar{b}$.

%\subsection{Event selection}
%\label{subsec:event_sel}
The event selection is developed in  order to reduce the background and increase the sensitivity to DM annihilations. An initial series of cuts is applied on to reduce the overall rate and to allow sophisticated but computationally expensive algorithms to run on a smaller number of high quality events.

\begin{figure}[p]
	\centering
	\includegraphics[width=1\textwidth]{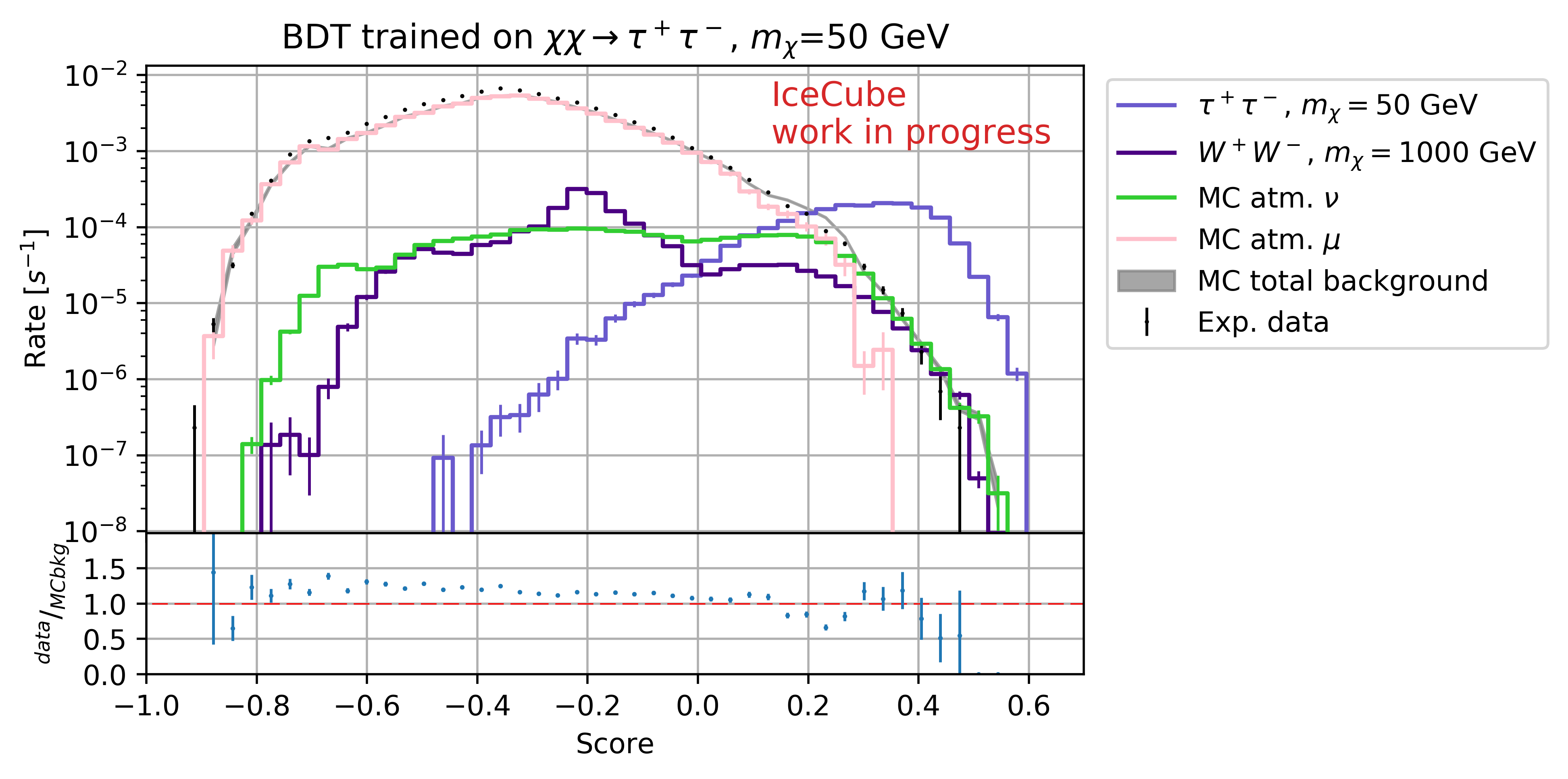}
	\caption{BDT Scores obtained with a BDT trained on $m_{\chi}=50$ GeV and annihilation channel $\chi\chi \rightarrow \tau^+\tau^-$. The violet lines indicate the two benchmark signal distributions. Distributions of atmospheric muons and neutrinos are shown in pink and green, respectively. Total MC background and experimental data are shown as a grey band and a dotted black line, respectively. Only statistical uncertainties are included.}
	\label{bdt11scores}
\end{figure}

\begin{figure}[p]
	\centering
	\includegraphics[width=1\textwidth]{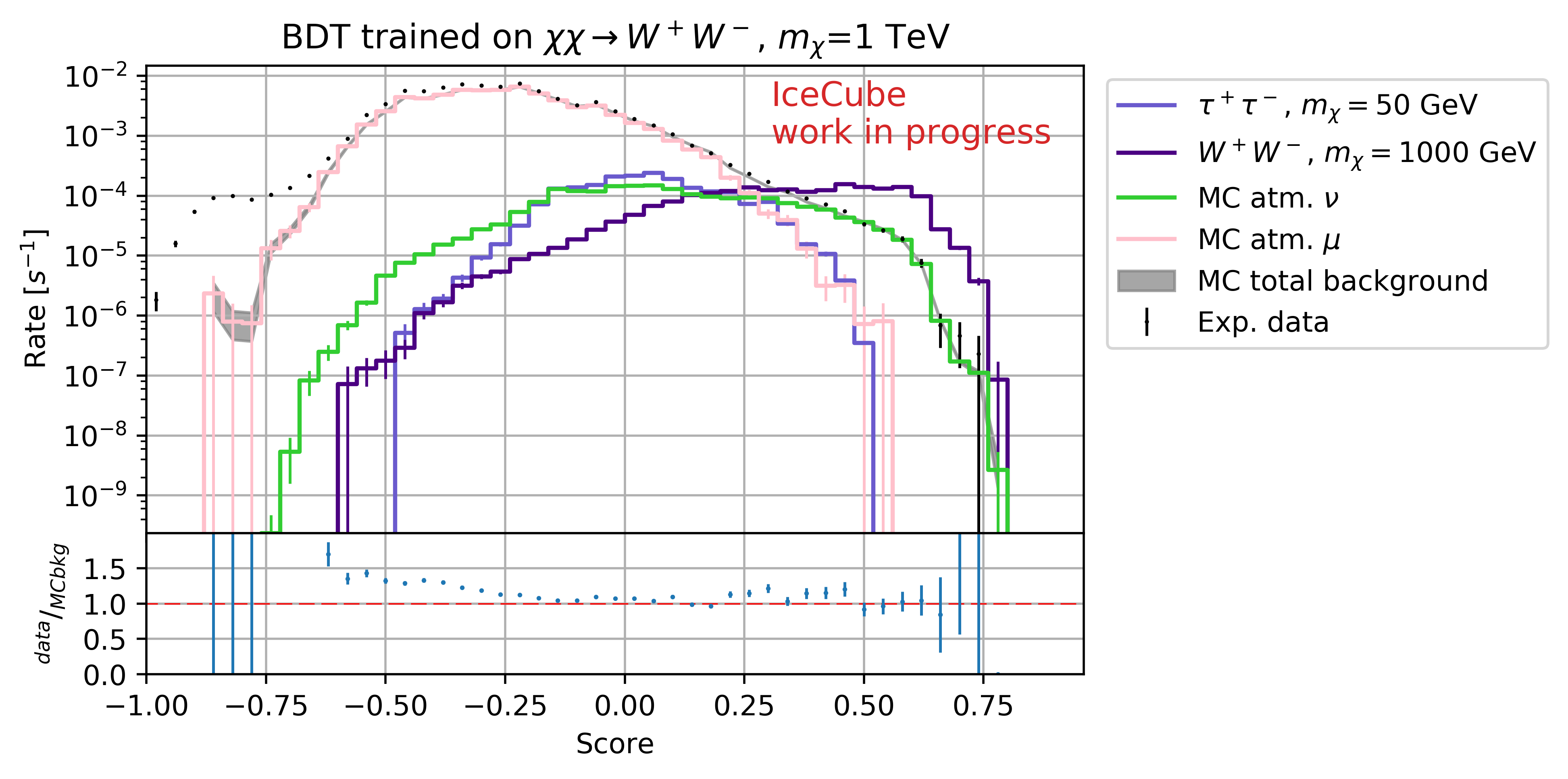}
	\caption{BDT Scores obtained with a BDT trained on $m_{\chi}=1$ TeV and annihilation channel $\chi\chi \rightarrow W^+W^-$. The violet lines indicate the two benchmark signal distributions. Distributions of atmospheric muons and neutrinos are shown in pink and green, respectively. Total MC background and experimental data are shown as a grey band and a dotted black line, respectively. Only statistical uncertainties are included.}
	\label{bdt8scores}
\end{figure}

The final stage of the selection process is a Boosted Decision Tree (BDT). This machine-learning procedure is capable of efficiently separating signal from background, assigning to each event a score that indicates how signal-like the event is. Two different BDTs were trained on the two benchmark signals mentioned above. In Figs. \ref{bdt11scores} and \ref{bdt8scores} the results for the two BDTs are shown. In both cases, the agreement between experimental data and MC improves for higher BDT scores. The final cut value on the BDT score is optimized in order to obtain the best possible sensitivity. After the cut is applied, the rate of background neutrinos is comparable to that of muons.

\section{Analysis method and sensitivity}
In order to obtain a sensitivity estimate on the annihilation rate  $\Gamma_A$, a statistical method similar to the one used in  \cite{EarthIce:2016} is applied. A binned likelihood test is performed. The likelihood to observe a certain number of events inside a given bin distribution is then defined as
\begin{equation}
\label{eq:binned_lh}
\mathcal{L}(\mu)=\prod_{\textrm{bin}_i=\textrm{bin}_{min}}^{\textrm{bin}_{max}}\mbox{Poisson}(N_{\rm obs}(\textrm{bin}_i)|N_{\rm obs}^{\rm tot}f(\textrm{bin}_i|\mu))\textrm{ ,}
\end{equation}
where $f(bin_i|\mu)$ is defined as fraction of total events falling inside the bin $i$:
\begin{equation}
f(\textrm{bin}_i|\mu)=\mu S(\textrm{bin}_i) + (1-\mu)B(\textrm{bin}_i)\textrm{ ,}
\label{eq:single_bin}
\end{equation}
with $\mu=\frac{N_{\rm sig}}{N_{\rm tot}}$ being the fraction of signal events contained in the data sample. The functions $S(\textrm{bin}_i)$ and $B(\textrm{bin}_i)$ are the probability density distributions (PDFs) describing signal and background zenith angle distribution respectively. Both are derived from simulations after the BDT cut is applied.

%Following the procedure described in \cite{FC:98}, the value $\mu_{\rm best}$ minimizing the likelihood is found in order to define a ranking function:

%\begin{equation}
%	\label{eq:fc_rank}
%	R(\mu)=\frac{\mathcal{L}(\mu)}{\mathcal{L}(\mu_{\rm best})}\textrm{ .}
%\end{equation}

%Values of $\mathcal{L}(\mu)$ (eq. \ref{eq:binned_lh}) are added following the ordering defined by eq. \ref{eq:fc_rank} until the desired 90\% confidence level (C.L.) is reached, defining the upper and lower limits. The lower limit is physically constrained to be larger than or equal to 0. The upper limit $\mu^{0.9}$ multiplied by the total number of events in the data sample gives the upper limit on the number of signal events $N_{\rm sig}^{0.9}$.

Using the Feldman-Cousins approach \cite{FC:98}, the 90\% upper limit on the number of signal events $N_{\rm sig}^{0.9}$ is found. DM annihilation muon neutrinos interact in the vicinity of the IceCube detector producing muons and defining a volumetric neutrino flux which is given by:
\begin{equation}
	\label{eq:conv_rate}
	\Gamma_{\nu\rightarrow\mu} = \frac{N_{\rm sig}^{0.9}}{t_{\rm live}\cdot V_{\rm eff}} \textrm{ ,}
\end{equation}
where $t_{\rm live}$ is the livetime of the experiment and $V_{\rm eff}$ is the effective volume of the detector. Finally, $\Gamma_{\nu\rightarrow\mu}$ values are converted to annihilation rates $\Gamma_A$ using WimpSim.
%This way, limits on the annihilation cross-section and on the spin-independent cross-section described in Section \ref{sec:theory} can then be derived.

\begin{figure}
	\centering
	\includegraphics[width=.8\textwidth]{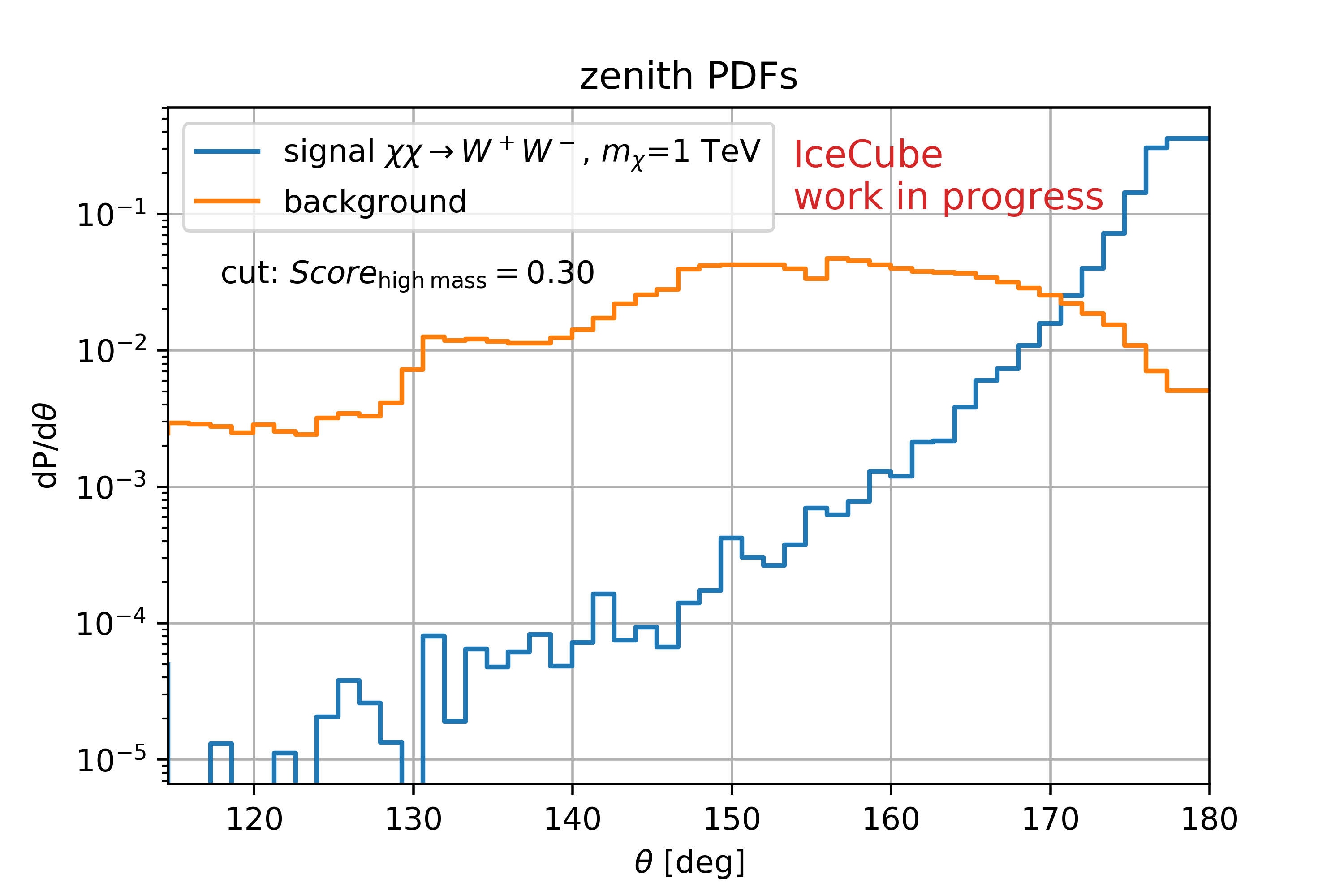}
	\caption{Background PDF (orange) and signal PDF (blue) for channel $\chi\chi\rightarrow W^+W^-$, $m_{\chi}=1$ TeV. Final high mass BDT cut value is $0.30$}
	\label{fig:pdfs}
\end{figure}

\begin{figure}
	\centering
	\includegraphics[width=.8\textwidth]{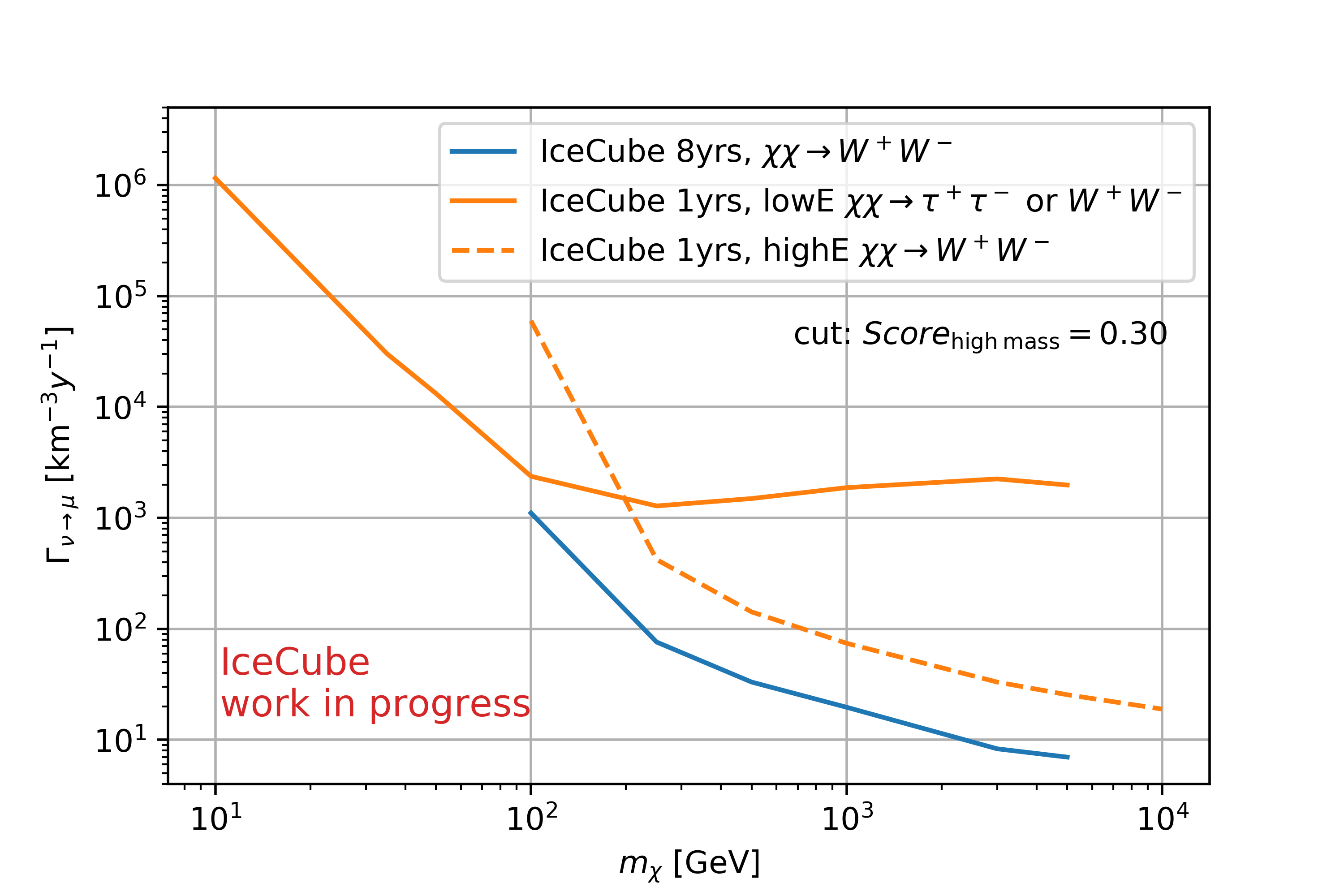}
	\caption{Estimated sensitivity (blue solid line) on the conversion rate for the annihilation channel $\chi\chi\rightarrow W^+W^-$ with a BDT cut of $0.30$, compared to the previous IceCube result \cite{EarthIce:2016} (orange).}
	\label{fig:sens_conv_rate}
\end{figure}

\begin{figure}
	\centering
	\includegraphics[width=.8\textwidth]{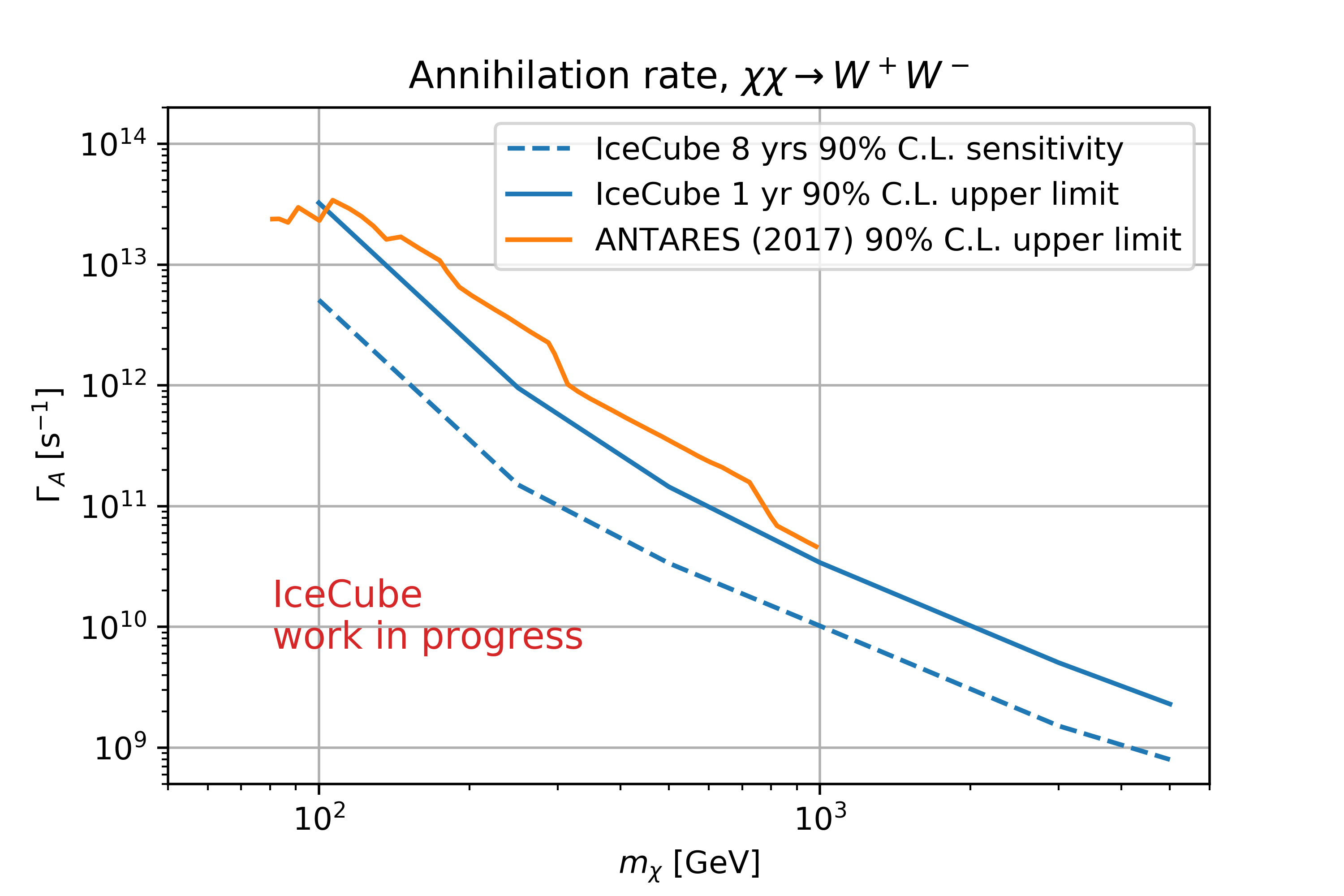}
	\caption{Sensitivity on the annihilation rate for the annihilation channel $\chi\chi\rightarrow W^+W^-$, compared with the upper limits from IceCube \cite{EarthIce:2016} (blue) and ANTARES \cite{ANTARES_Earth:2016} (orange).}
	\label{fig:ann_rate}
\end{figure}

Sensitivity is defined as the median upper limit $\widetilde{N}_{\rm sig}^{0.9}$ obtained from $10^4$ pseudo-experiments that were built under the hypothesis of no signal (only background). Sensitivities were computed for the annihilation channel $\chi\chi\rightarrow W^+W^-$, using the \textit{high mass} trained BDT. A preliminary scan on the BDT score cut identified the region around $0.3$ as the optimum cut for the best sensitivity. The corresponding signal and background PDFs, built with the selected BDT cut, are shown in Fig. \ref{fig:pdfs}. The experimental rate after the cut is $0.19$ mHz, corresponding to a total number of $N_{\rm obs}^{\rm tot}=43032$ events for the full considered livetime. In Figs. \ref{fig:sens_conv_rate} and \ref{fig:ann_rate}, the calculated sensitivities are shown. The result on the volumetric flux is exceeding the previous IceCube sensitivity \cite{EarthIce:2016} by a factor of $\sim3.8$ for $m_{\chi}=1$ TeV. The sensitivity on the annihilation rate exhibits a significant improvement when compared to the previous limits from IceCube and ANTARES \cite{ANTARES_Earth:2016}.

A work to improve the previously presented analysis method is currently being developed. Firstly, the likelihood can be calculated on an event-by-event basis. Secondly, the likelihood can be extended to include the reconstructed energy of the events as a parameter in addition to the zenith angle. Both changes add more information to the likelihood, increasing the potential of the method to obtain even better sensitivities. Every event is then described by the 2-D vector $\vec{\xi}=(\theta,E_{\rm reco})$, for which we can define the probability to be observed:
\begin{equation}
f(\vec{\xi}|\mu)=\mu S(\vec{\xi}) + (1-\mu)B(\vec{\xi})\textrm{ ,}
\label{eq:single_event}
\end{equation}
where the functions $S(\vec{\xi})$ and $B(\vec{\xi})$ are now event-wise.

The likelihood to observe $N_{\rm tot}$ events given a certain signal fraction $\mu$ is then defined as:
\begin{equation}
\label{eq:likelihood}
\mathcal{L}(\mu)=\prod_{i=0}^{N_{tot}}f(\vec{\xi}_i|\mu)\textrm{ .}
\end{equation}

Limits and sensitivity can be calculated as described before, using the new likelihood definition.

\section{Conclusion}
A first sensitivity study has been performed for the on-going analysis searching for DM annihilation in the center of the Earth with 8 years of IceCube data. A new event selection has been developed and the likelihood method used in the previous search by IceCube has been tested. The results obtained show a significant improvement that could lead to world competitive limits on the spin-independent dark matter-nucleon scattering cross-section. Further improvements are expected by updating the analysis method to an event-wise unbinned likelihood and by extending the method to include the reconstructed energy as a parameter in the likelihood.

%\newpage

%\section{Listing some References}\label{sec:refs}

%This is a paper from a previous ICRC \cite{Zoll:2015wcu}. This is a second paper from a previous ICRC \cite{Peiffer:2017vsm}. This is a paper from the current ICRC \cite{Hussain:2019icrc_gw}.
%Here is an IceCube journal paper \cite{Aartsen:2016nxy} and an external journal paper \cite{Waxman:1998yy}.

% Set up the bibliography using BibTeX.
% Get references from inspirehep.net or NASA/ADS and put them in references.bib.
\bibliographystyle{ICRC}
\bibliography{references}

% Or, set up the bibliography manually, if you prefer to do things this way.
%
% \begin{thebibliography}{99}
%   \bibitem{Zoll:2015wcu}{{\bf IceCube} Collaboration, \pos{PoS(ICRC2015)1099} (2016).}
%   \bibitem{Peiffer:2017vsm}{{\bf IceCube-Gen2} Collaboration, \pos{PoS(ICRC2017)1052} (2018).}
%   \bibitem{Hussain:2019icrc_gw}{{\bf IceCube} Collaboration, \pos{PoS(ICRC2019)xyz} (these proceedings).}
%   \bibitem{Aartsen:2016nxy}{{\bf IceCube} Collaboration, M.~G.~Aartsen {et al.}, \emph{JINST} {\bf 12} (2017) P03012%
%   % optionally add arXiv ID here [{\tt astro-ph/1612.05093}]
%   .}
%   \bibitem{Waxman:1998yy}{E. Waxman and J. N. Bahcall, \emph{Phys. Rev.} {\bf D59} (1999) 023002.}
% \end{thebibliography}

\end{document}